\documentclass[twocolumn,preprintnumbers,amsmath,prd,amssymb]{revtex4}

\usepackage{graphicx}% Include figure files
\usepackage{dcolumn}% Align table columns on decimal point
\usepackage{bm}% bold math
\usepackage{amsmath}
\usepackage{lipsum}
\usepackage{slashed}
\usepackage{float}
\usepackage{xcolor}

\begin{document}

\title{Neutron stars: New constraints on asymmetric dark matter}

\author{O. Ivanytskyi}
\email{oivanytskyi@usal.es}
\affiliation{CFisUC, Department of Physics, University of Coimbra, Rua Larga P-3004-516, Coimbra, Portugal}
\affiliation{Department of Fundamental Physics, University of Salamanca, Plaza de la Merced S/N E-37008, Salamanca, Spain}
\affiliation{Bogolyubov Institute for Theoretical Physics, Metrologichna str. 14$^B$, Kyiv 03680, Ukraine}

\author{V. Sagun}
\email{violetta.sagun@uc.pt}
\affiliation{CFisUC, Department of Physics, University of Coimbra, Rua Larga P-3004-516, Coimbra, Portugal}

\author{I. Lopes}
\email{ilidio.lopes@tecnico.ulisboa.pt}
\affiliation{Centro de Astrof\'{\i}sica e Gravita\c c\~ao  - CENTRA, Departamento de F\'{\i}sica, Instituto Superior T\'ecnico - IST, Universidade de Lisboa - UL, Av. Rovisco Pais 1, 1049-001 Lisboa, Portugal}

%%%%%%%%%%%%%%%%%%%%%%%%%%%%%%%%%%%%%%%%%%%%%%%%%%%%%%%%%%%%%%%%
\date{\today}

\begin{abstract} 
We study an impact of asymmetric dark matter on properties of the neutron stars and their ability to reach the two solar masses limit, which allows us to present a new range of masses of dark matter particles and their fractions inside the star. Our analysis is based on the observational fact of the existence of two pulsars reaching this limit and on the theoretically predicted reduction of the neutron star maximal mass caused by the accumulation of dark matter in its interior. We also demonstrate that light dark matter particles with masses below 0.2 GeV can create an extended halo around the neutron star leading not to decrease, but to increase of its visible gravitational mass. By using recent results on the spatial distribution of dark matter in the Milky Way, we present an estimate of its fraction inside the neutron stars located in the Galaxy center. We show how the detection of a $2 M_\odot$ neutron star in the most central region of the Galaxy will impose an upper constraint on the mass of dark matter particles of $\sim$ 60 GeV. Future high precision measurements of the neutron stars maximal mass near the Galactic center, will put a more stringent constraint on the mass of the dark matter particle. This last result is particularly important to prepare ongoing, and future radio and X-ray surveys.

\end{abstract}

%\pacs{ln}

\maketitle %must follow title, authors, abstract, \pacs, and \keywords
  
%%%%%%%%%%%%%%%%%%%%%%%%%%%%%%%%%%%%%%%%%%%%%%%%%%%%%%%%%%%
\section{Introduction} 

Despite strong observational evidences of dark matter (DM) existence \citep{2016A&A...594A..13P}, at present, its fundamental nature remains a mystery, which is the reason why there  exist so many DM candidates. Among the most promising ones are WIMPs, axions \citep{2000PhR...325....1R} and sterile neutrinos \citep{1994PhRvL..72...17D}. Unfortunately, until now, terrestrial experiments on nuclear recoil of DM \citep{2016PhR...627....1M}, and direct searches of the DM annihilation \citep{2014arXiv1411.1925C} have not yet found a suitable candidate. 

\paragraph*{}
Because of it, the astrophysical probes of the DM properties are of highest interest. Compact astrophysical objects, such as neutron stars (NSs) are especially attractive in this context, since they can accumulate a sizable amount of DM in the stellar interior \citep{2010PhRvD..82f3531K,2013PhRvD..88l3505B,2006PhRvD..74f3003N,2017PhRvD..96b3002P}. At the same time, depending on its nature, DM can affect the properties of NSs in quite different ways. For instance, an accretion of self-annihilating DM to a NS will increase its luminosity and effective temperature \citep{2012PhLB..711....6P}.
In addition, as was shown in the recent studies \citep{2019PhRvD.100d4049B, 2018PhLB..781..607E} presence of the DM inside the NSs during their merger might produce a distinguishable signature in the GW signal, especially during the post-merger stage.

\paragraph*{}
A challenging new way to study DM  is to consider that the DM particles are similar to those of baryon matter (BM). Likewise, this new particle caries a conserved charge, and, as a consequence, causes an absence of symmetry between such particles and antiparticles \citep{2009JCAP...07..004F,2014ApJ...780L..15L}.  Hence, due to the absence of anti-particles, no self-annihilation can occur, and as such, this process produces no heat. Such type of DM is usually called asymmetric dark matter (ADM). There is a large number of articles that have explored ADM in different astrophysical contexts. Examples are the studies of ADM  in the Sun~\citep{2012ApJ...757..130L,2015PhRvL.114h1302V,2016JCAP...11..007V,2016PhRvD..94f3512M} and other stars~\citep{2017PhRvD..95b3507M,2019ApJ...879...50L}.
This DM type is the subject of our study in this manuscript.

\paragraph*{}
Several robust observational results strongly support the existence of DM \citep{2017FrPhy..12l1201Y}.  For instance, in Ref. \cite{2019A&A...621A..56P} is reported that in the Galaxy, about 70\% of its total mass is DM. Therefore, if ADM exists, in principle, its fraction inside the NSs can be significant. In the absence of repulsive self-interactions, the gravitational collapse of ADM can lead to the formation of a mini black hole inside the NS; this imposes a powerful constraint on the properties of DM \citep{2012PhRvL.108s1301K}. If, however, ADM is constituted by fermions or bosons with self-repulsion, then it can resist gravitational collapse and form stable configurations inside the NSs \citep{2013PhRvD..87l3507B}. 

\paragraph*{}
The main consequence resulting from the presence of DM inside the NSs is the reduction of the stellar mass (see 
\citep{2018PhRvD..97l3007E, 2019PhRvD..99f3015D} and references therein). This measurable effect provides an opportunity to infer the DM properties from the mass-radius relation for NSs. \cite{arXiv:1904.13060} has shown that the mass of pulsars admixed with DM depends on its distance from the Galaxy center. Presently, however, this approach is limited, since it requires reliable observational data on many compact objects. At the same time, a significant reduction of the NS mass caused by the presence of DM, can make the two solar mass limit unattainable and shift the maximal apparent mass to lower values. Thus, observational fact of the existence of two NSs that have the largest known masses (i.e., PSR J0348+0432, PSR J0740+6620) \cite{2013Sci...340..448A, 2019arXiv190406759C}, enables us to formulate a strict constraint on the maximum mass of an ADM particle, which is the primary goal of this study.

In Sec. \ref{sec-2} we present the equations of state (EoS) of DM and BM. This section is also devoted to the system of TOV-like equations for baryon and dark components, which derivation is given in Appendix. In Sec. \ref{sec-3} an effect of DM on the NS properties for different masses of DM particles and fractions inside the NSs is discussed. In Sec. \ref{sec-4} we estimate the amount of DM in the neighbourhood of the two heaviest pulsars. Secs. \ref{sec-5} and \ref{sec-6} are devoted to accretion of DM onto NS at the different stages of its evolution and to the observational signatures of the DM admixed NSs, respectively. Our conclusions follow in Sec. \ref{sec-7}.
 
%%%%%%%%%%%%%%%%%%%%%%%%%%%%%%%%%%%%%%%%%%%%%%%%%%%%%%%%%%%
\section{Dark matter admixed neutron stars}
\label{sec-2}

The current experimental bounds for the DM direct detection imposes the DM-BM interaction cross-section $\sigma_{D}\sim10^{-45}$ cm$^2$ \citep{2016JPhG...43a3001M} to be many orders of magnitude lower than a typical nuclear one $\sigma_N\sim10^{-24}$ cm$^2$. Conveniently, following the common logic \cite{2017PhRvD..96b3002P,Sandin2009,Sandin2011} we neglect the interaction between baryons and DM particles in the rest of the article. Accordingly, we assume that the BM and DM are coupled only through gravity, and their energy-momentum tensors are conserved separately. Hence, the system of equations 
\begin{eqnarray}
\label{I}
\frac{dp_j}{dr}=-\frac{(\epsilon_j +p_j)(M+4\pi r^3p)}{r^2\left(1-{2M}/{r}\right)}
\end{eqnarray}
describes the relativistic hydrostatic equilibrium of a NS with DM. Hereafter subscript index $j=B$ and $j=D$ stands for BM and DM, respectively; $p_j$ and $\epsilon_j$ correspond to the pressure and energy density of the $j$ component, $r$ is the distance from the centre of the star, and $p(r)=p_B(r) +p_D(r)$. The gravitational mass $M$ is the sum of masses of both components, i.e., $M(r)=M_{B} (r)+M_{D}(r)$, where
\begin{eqnarray}
\label{II}
M_j (r)=4\pi\int\limits_0^r \epsilon_j (r') {r'}^2 dr'.
\end{eqnarray}
The system of Eqs. (\ref{I}) along with a definition of $M$ is a generalization of the Tolman-Oppenheimer-Volkov equation (TOV) \citep{1939PhRv...55..364T, 1939PhRv...55..374O}, recovered in the absence of the DM, when $p_D=0$ and $\epsilon_D=0$. This system directly comes from the Einstein's field equations and allows beyond the perturbative description of the relativistic hydrostatic problem with an arbitrary number of components, which interact only through gravity \cite{Sandin2009,Sandin2011}. For the readers convenience we present the corresponding derivation in Appendix.

We compute the radial density profiles of the BM and DM from Eqs. (\ref{I}) and (\ref{II}), by imposing two conditions at the centre of the star and two boundary conditions. The first set corresponds to the central densities of BM and DM. The second one corresponds to requiring the hydrostatic equilibrium of each of the matter components on their boundaries being spheres of radii $R_B$ and $R_D$, which contain all the BM and DM, respectively. Note, that $R_B\neq R_D$ in the general case. Such a hydrostatic equilibrium is provided independently for BM and DM by two conditions
\begin{eqnarray}
\label{III}
p_j (R_j)=0.
\end{eqnarray}

The DM component is not accessible by direct observations. Hence, it is natural to identify the NS radius $R$ with $R_B$, i.e. $R=R_B$. The total gravitational mass, and the fraction of DM inside the NS are defined as
\begin{eqnarray}
\label{IV}
M_{T}&=&M_B(R_B)+M_D(R_D),\\
\label{V}
f_\chi&=&\frac{M_{D}(R_D)}{M_{T}},
\end{eqnarray}
respectively. It is clear that variation of central densities of BM and DM allows us to obtain different values of $M_{T}$ and $R$ at given $f_{\chi}$. 

\subsection{BM equation of state} 

The reliable modelling of the DM distribution inside NSs requires all the effects of BM to be taken under control. For this purpose we use the recently developed realistic EoS with induced surface tension (IST EoS hereafter) \citep{2014NuPhA.924...24S, 2019ApJ...871..157S}, which accounts for the short-range repulsion between baryons, and their long-range attraction of the mean field type. This EoS is consistent with nuclear and hadron matter experimental data, i.e. the nuclear matter ground state properties \citep{2014NuPhA.924...24S}, the proton flow data \citep{2018PhRvC..97f4905I}, constraints on the hadron hard-core radii obtained in the heavy-ion collisions \citep{2018EPJA...54..100S, 2017EPJWC.13709007S}. Furthermore, recently, the IST EoS was successfully applied to modelling of purely baryon-lepton NSs \citep{2019ApJ...871..157S}. In this work, we stay with a parametrization of the present EoS found in Ref. \cite{2019Sagun}. 

\paragraph*{}
Partial derivatives of $p_B$ with respect to chemical potentials of neutrons $\mu_n$, protons $\mu_p$ and electrons $\mu_e$ give the number densities of corresponding particles, i.e. $n_n$, $n_p$ and $n_e$. Note that the baryonic charge density is $n_B=n_n+n_p$. Conditions of electric neutrality $n_p=n_e$ and equilibrium concerning $\beta$-decay $\mu_n=\mu_p+\mu_e$ exclude $\mu_p$ and $\mu_e$ from the list of independent variables. Thus, $p_B$ is a function of $\mu_n$ only, which coincides with the chemical potential associated with the baryonic charge $\mu_B$, i.e. $\mu_n=\mu_B$. The energy density of BM is defined through the well known thermodynamic identity as
\begin{eqnarray}
\label{VI}
\epsilon_B =\sum_{i=n,p,e}\mu_i n_i-p_B =\mu_B n_B-p_B .
\end{eqnarray}

\subsection{DM equation of state} 

Here we consider the DM component as a relativistic Fermi gas of noninteracting particles with the spin one-half. The corresponding EoS has been exhaustively described in the literature, for instance in Ref. \cite{2018arXiv180303266N}. The energy density of DM is obtained from the thermodynamic identity
\begin{eqnarray}
\label{VII}
\epsilon_D=\mu_D n_D-p_D,
\end{eqnarray}
where $\mu_D$ and $n_D={\partial p_D}/{\partial \mu_D}$ are chemical potential and particle number density of DM. Typically, its repulsive self-interaction is accounted for by the term ${n_D^2}/{m_I^2}$ in pressure, where $m_I$ is an interaction scale. We neglect this term, since, for example, at $m_I=m_\chi$ it does not exceed 15 \% of $p_D$ even in the NS centre, where $n_D$ reaches the highest value.

\subsection{Relation between chemical potentials}

An important feature of the solution of system of Eqs. (\ref{I}) can be drawn without its numerical analysis. It corresponds to the relation between chemical potentials of BM and DM. In order to derive such relation it is convenient to use the thermodynamic identity $n_j=\frac{\partial p_j}{\partial\mu_j}$, which yields $\frac{dp_j}{dr}=n_j\frac{d\mu_j}{dr}$. This relation along with Eqs. (\ref{VI})-(\ref{VII}) allows us to rewrite the system of Eqs. (\ref{I}) in the form 
\begin{eqnarray}
\label{VIII}
\frac{d\ln\mu_B }{dr}=\frac{d\ln\mu_D}{dr}=-\frac{M+4\pi r^3p}{r^2\left(1-{2M}/{r}\right)}.
\end{eqnarray}
From this result, we immediately conclude that chemical potentials $\mu_B$ and $\mu_D$ are proportional to each other. Their values in the centre of a NS define the proportionality coefficient, i.e.
\begin{eqnarray}
\label{IX}
\mu_D=\mu_B\cdot\frac{\mu_D}{\mu_B}\Bigl|_{r=0}
\end{eqnarray}
This conclusion is in line with the fact that in the equilibrium case chemical potential defined in local Lorentz frame scales as $\mu_j\sqrt{g_{44}}=const$ (see e.g. \cite{2018arXiv180303266N}), which leads to $\frac{\mu_D}{\mu_B}=const$.

%%%%%%%%%%%%%%%%%%%%%%%%%%%%%%%%%%%%%%%%%%%%%%%%%%%%%%%%%%%
\section{Gravitational influence of DM condensate}  
\label{sec-3}

\begin{figure}[!]
\centering
\includegraphics[scale=0.55]{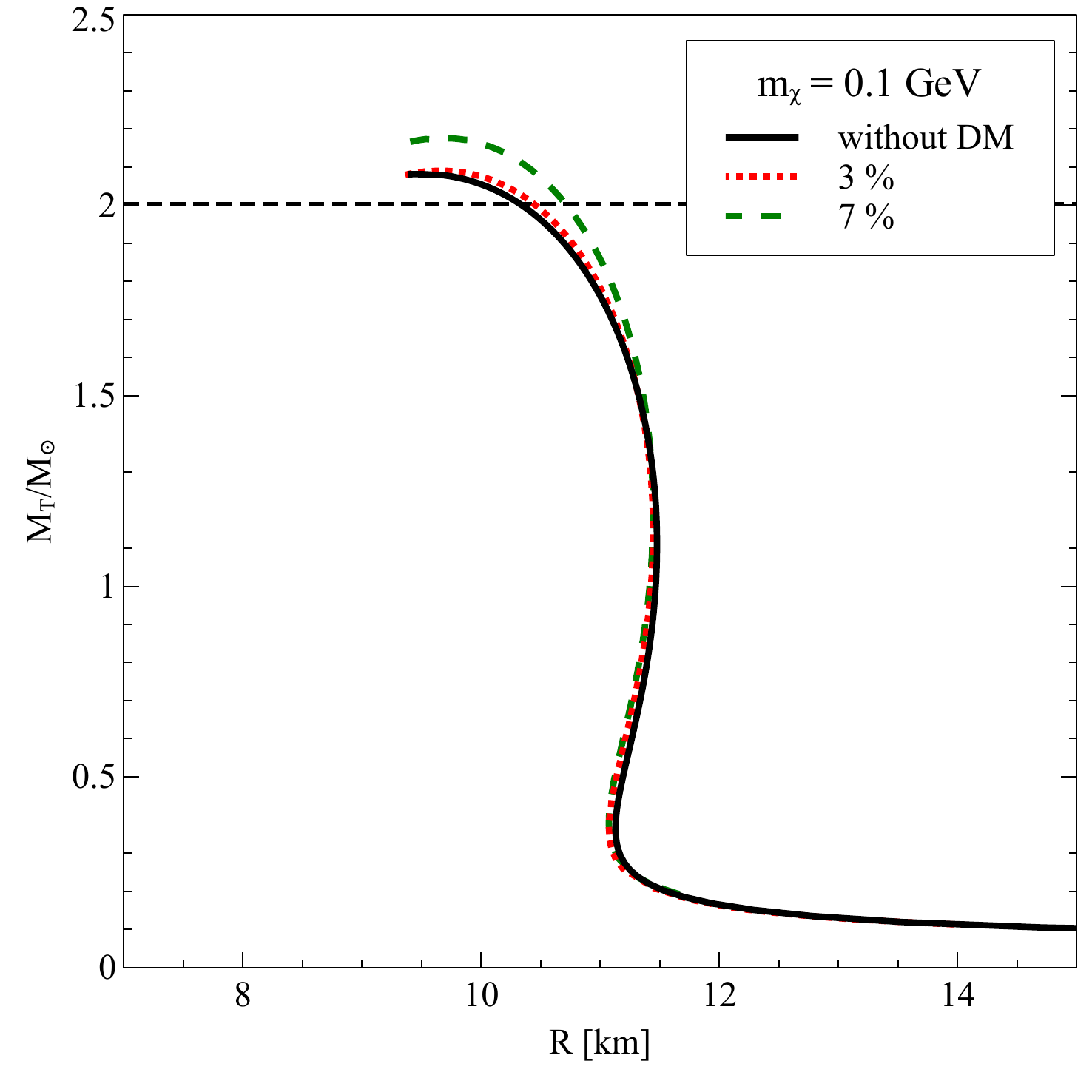}
\includegraphics[scale=0.55]{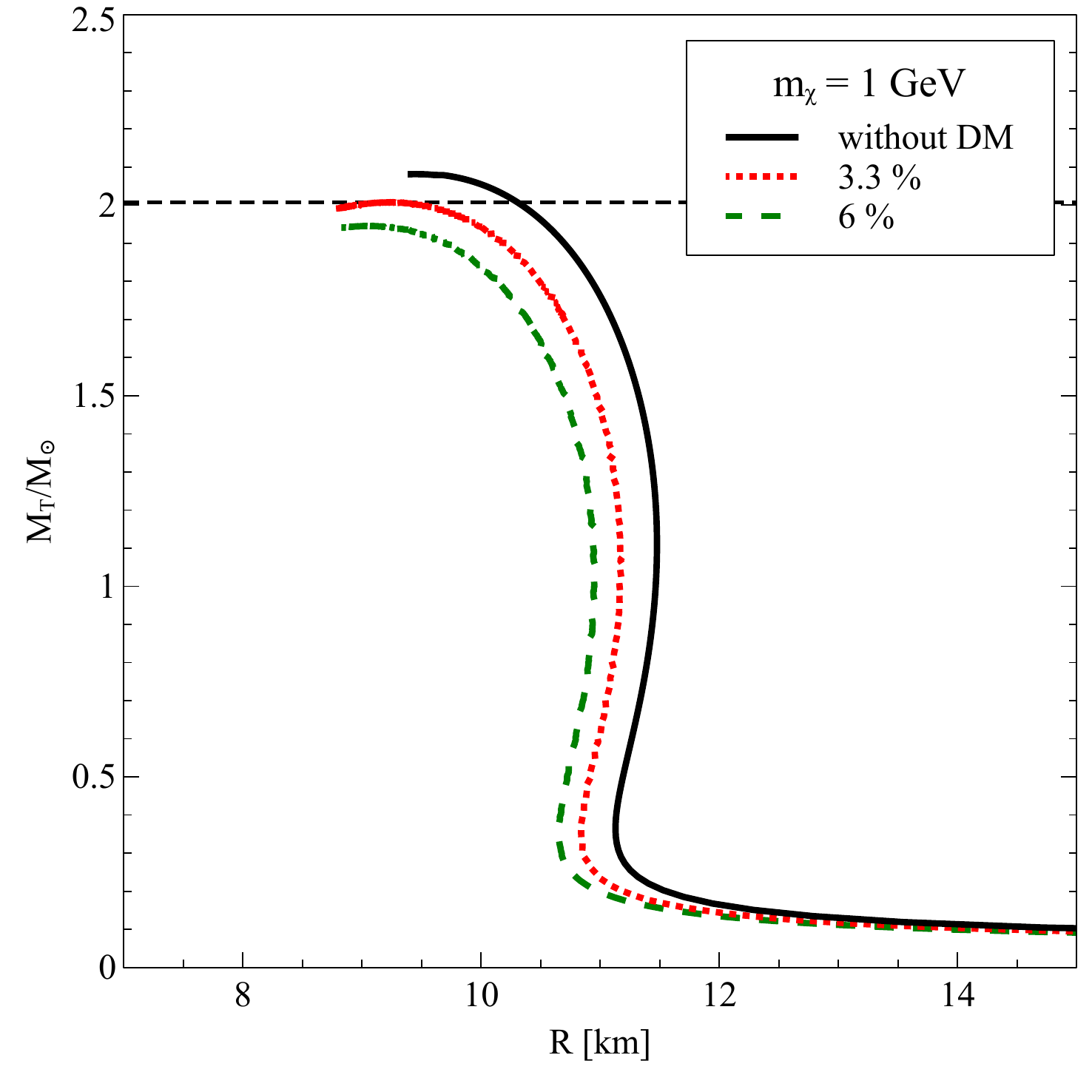}
\caption{Total gravitational mass $M_{T}$ of the DM admixed NS vs. its visible radius $R$ calculated for $m_\chi=$ 0.1 GeV (upper panel) and 1 GeV (lower panel) for the different fractions of DM $f_\chi$. The straight dashed line corresponds to $M=2~M_\odot$.}
\label{fig1}
\end{figure}

\begin{figure}[!]
\includegraphics[scale=0.55]{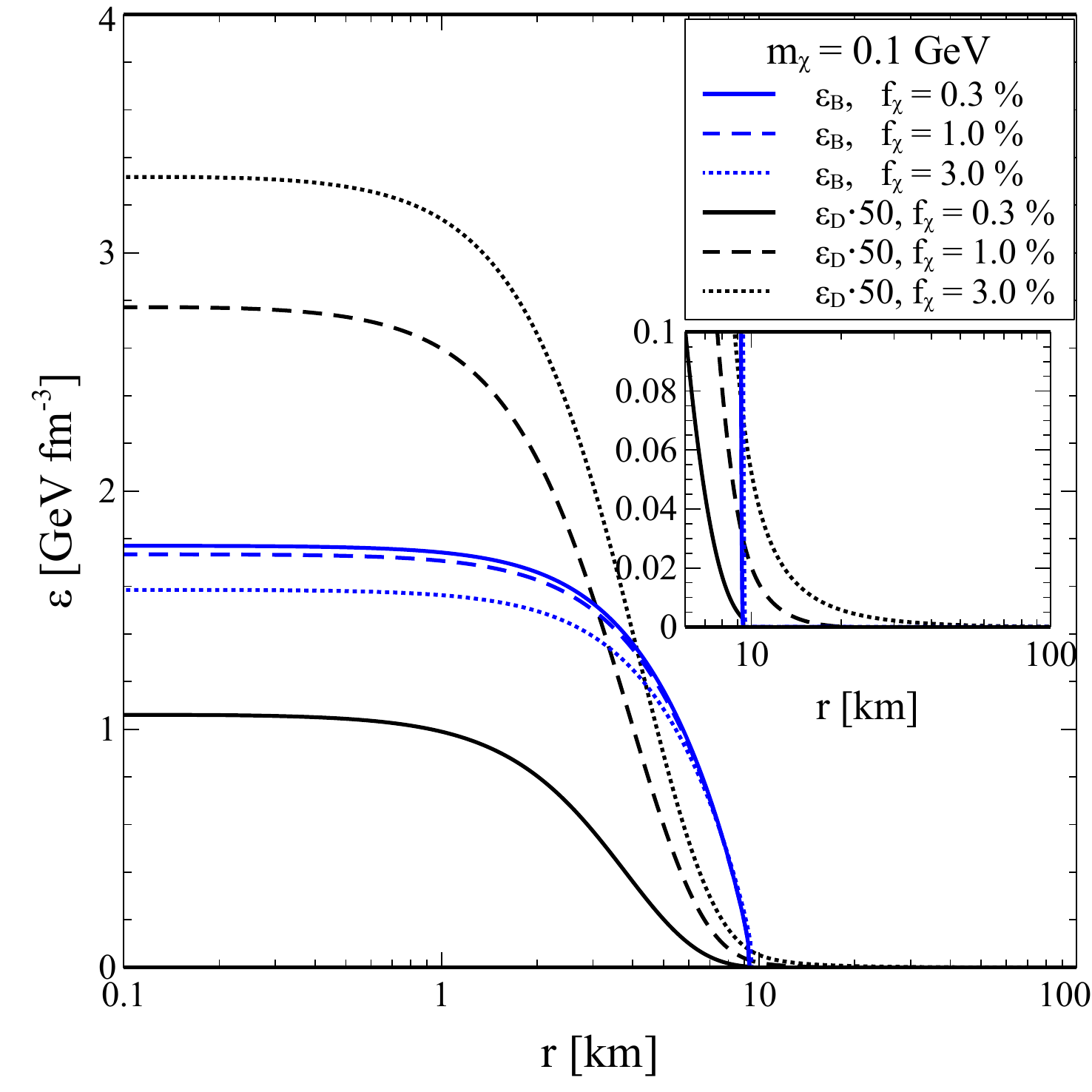} 
\includegraphics[scale=0.55]{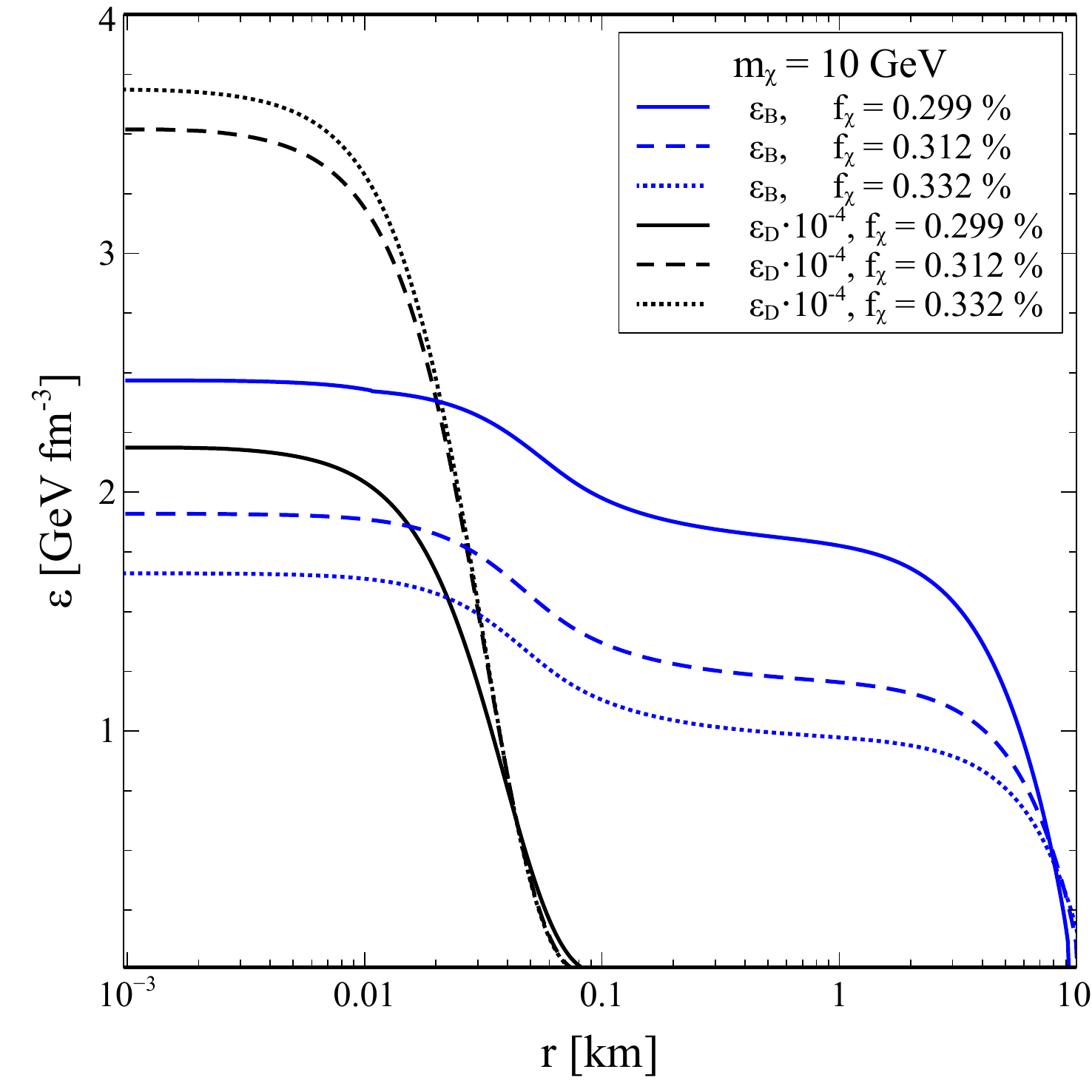} 
\caption{Energy density profiles of BM (blue curves) and DM (black curves) inside NS of the maximal mass $M_T=M_{max}$, that can be reached at a given value of $f_\chi$. Calculations are performed for $m_\chi = 0.1~GeV$ (upper panel) and $m_\chi = 10~GeV$ (lower panel). To facilitate the reading $\epsilon_D$ is scaled by factors of 50 (upper panel) and $10^{-4}$ (lower panel).}
\label{fig2}
\end{figure}

Once we solved a two-component system of  Eqs.  (\ref{I})  with the appropriate boundary conditions, we can analyse the influence of the DM on the total gravitational mass of a NS and its visible radius. Fig. \ref{fig1} shows the mass-radius relation of the DM admixed NS for two values of $m_\chi$ and different fractions $f_\chi$. We found that light DM particles can not sizably reduce the maximal mass of NS $M_{max}$. Moreover, we note that $m_\chi=$ 0.1 GeV yields $M_{max}>2~M_\odot$ for any $f_\chi$. However, for heavy DM particles the situation changes completely. As it is seen from the lower panel of Fig. \ref{fig1}, DM with $m_\chi=$ 1 GeV strongly influences the mass-radius relation of NSs. For example, for $f_\chi=$ 3.3 \% (red dotted curve) $M_{max}$ equals to $2~M_\odot$, while further increase of the DM fraction leads to decrease of $M_{max}$ below $2~M_\odot$ (green dashed curve). For larger  values of $m_\chi$ the reduction of $M_{max}$ is even more dramatic. Such a sensitivity of the NSs mass to the presence of DM is related to its distribution in the stellar interior. 

\paragraph*{}
Fig. \ref{fig2} shows the radial profiles of the energy density of BM and DM inside a NS. In the case of light DM particles (with $m_\chi$ = 0.1 GeV) these profiles are rather gradual at all fractions $f_\chi$. Besides that, the typical values of $\epsilon_D$ are significantly smaller than $\epsilon_B$, while $R_D$ is large. In particular, $R_D= 9.4$ km for $f_\chi = 0.3 \%$, $R_D = 21.2$ km for  $f_\chi = 1.0~\% $ and $R_D = 135.2$ km for $f_\chi = 3.0~\%$. Note, that these large values of $R_D$ relate to the existence of dilute and extended halos of DM around a baryon core of NS. As a consequence, at small $m_\chi$ DM does not form a compact structure inside the NS able to significantly reduce its total mass. This explains qualitatively why the presence of light DM particles inside a NS is consistent with  $M_{T}\ge2~M_\odot$. The situation is completely different in the case of heavy DM particles. The lower panel of Fig. \ref{fig2} demonstrates that the energy density profile of DM with $m_\chi$ = 10 GeV is very steep. Indeed, $\epsilon_D$  drops from its maximal value, being four orders of magnitude above $\epsilon_B$, to almost zero within only about $R_D$ = 0.1 km. Such a high value of $\epsilon_D$ in combination with small $R_D$ leads to a very compact core of DM. The corresponding compactness $2M_{D}(R_D)/{R_D}$ can reach values up to 0.4, while total compactness is even slightly larger. It follows from Eq.  (\ref{I}) that in this case the derivatives ${dp_B}/{dr}$ and ${dp_D}/{dr}$ become large by an absolute value. Hence, $\epsilon_B$ and $\epsilon_D$ rapidly decrease at $r\sim R_D$, which is seen on the lower panel of Fig. \ref{fig2}.

\begin{figure}[!]
\includegraphics[scale=0.55]{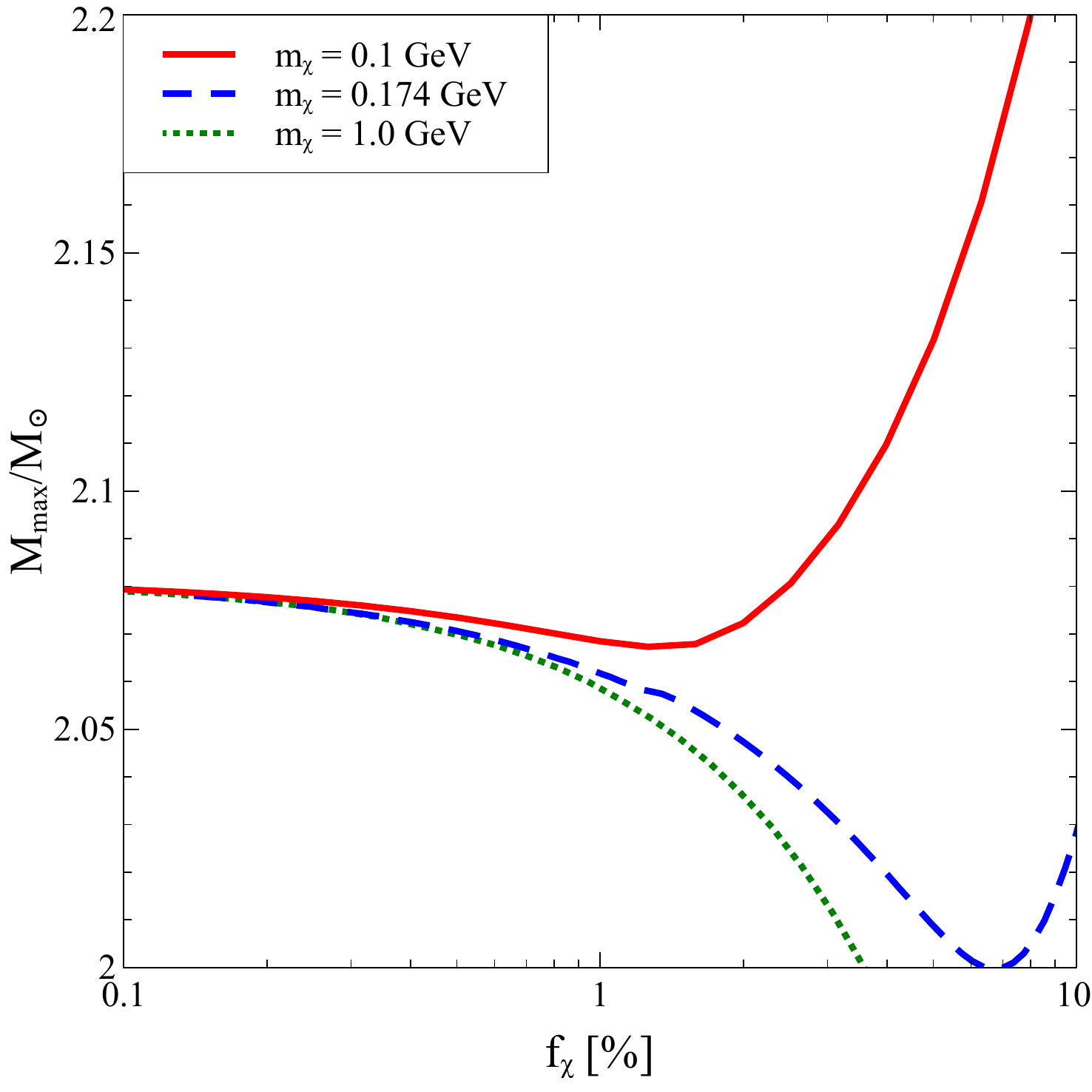}  
\caption{Maximal mass of NS $M_{max}$ as a function of the DM fraction $f_\chi$ at $m_\chi$ = 0.1 GeV (red solid curve), $m_\chi$ = 0.174 GeV (blue dashed curve) and $m_\chi$ = 1 GeV (green dotted curve).}
\label{fig3}
\end{figure}

At $r>R_D$ the impact of the DM core weakens and profile of the BM energy density becomes gradual again. At the same time, the values of  $\epsilon_B$ in the regions of the star with $r \sim$ 0.1 - 9 km  are considerably smaller than without DM. This leads to a significant reduction of the total mass of NS. Thus, we conclude that heavy DM particles tend to create very compact core, which even despite small fraction $f_\chi$, reduces the total mass of NS, and does not allow it to reach the two solar masses limit. As is seen from the lower panel of Fig. \ref{fig3}, such cores are characterized by very high energy densities, which is caused by rather weak ability of heavy DM to resist the gravitational compression. Such a resistivity can be quantified by the ratio $\frac{p_D}{\epsilon_D}$. At $m_\chi=0.1$ GeV this ratio is close to 0.25 already at $\epsilon_D=10^{-2}$ GeV fm${}^{-3}$, while for $m_\chi=$ 10 GeV it is below 0.01 even at $\epsilon_D=10^3$ GeV fm${}^{-3}$. This means that contrawise the light DM, heavy DM is not able to resist the gravitational compression until its energy density reaches even higher densities corresponding to compact cores. It is also worth noting that for large $m_\chi$, the energy density profiles inside NS are very sensitive to the fraction of DM. For example, at $m_\chi$  = 10 GeV a relative increase of $f_\chi$  by 7.6 \% (from 0.299 \% to 0.322 \%) leads to about 40 \% reduction of $\epsilon_B$ and, consequently, to a drastic decrease of the NS maximal mass from $2.07~M_\odot$ to $1.88~M_\odot$. 

\paragraph*{}
Thus, we conclude that light DM particles do not modify the mass-radius relation significantly, while heavier DM particles lead to a substantial reduction of the NS maximal mass.

\paragraph*{}
Fig. \ref{fig3} shows $M_{max}$ as a function of $f_\chi$ for several values of $m_\chi$.  As shown, for $m_\chi$ = 0.1 GeV (solid red curve), the most significant  reduction of the NS maximal mass is achieved at $f_\chi\simeq$ 1.3 \% with $M_{max} = 2.07~M_\odot$.We notice that the existence of an extended DM halo increases the NS maximal mass at larger fractions $f_\chi$. At the same time, for $m_\chi$ = 1.0 GeV (green dotted curve) $M_{max}$ monotonously decreases with the grows of $f_\chi$, and gets smaller than $2M_\odot$ already at $f_\chi=0.312~\%$. For $m_\chi$  = 0.174 GeV the lowest value of  $M_{max}$ is $2~M_\odot$. Therefore, within the present model, DM particles with $m_\chi\le$ 0.174 GeV are consistent with the $2~M_\odot$ constraint for any $f_\chi$, while for heavier DM particles the NS mass can reach $2~M_\odot$ only if $f_\chi$ is limited from above.

%%%%%%%%%%%%%%%%%%%%%%%%%%%%%%%%%%%%%%%%%%%%%%%%%%%%%%%%%%%
\section{Constraining mass of the dark matter particles}
\label{sec-4} 

\begin{figure}[!]
\includegraphics[scale=0.55]{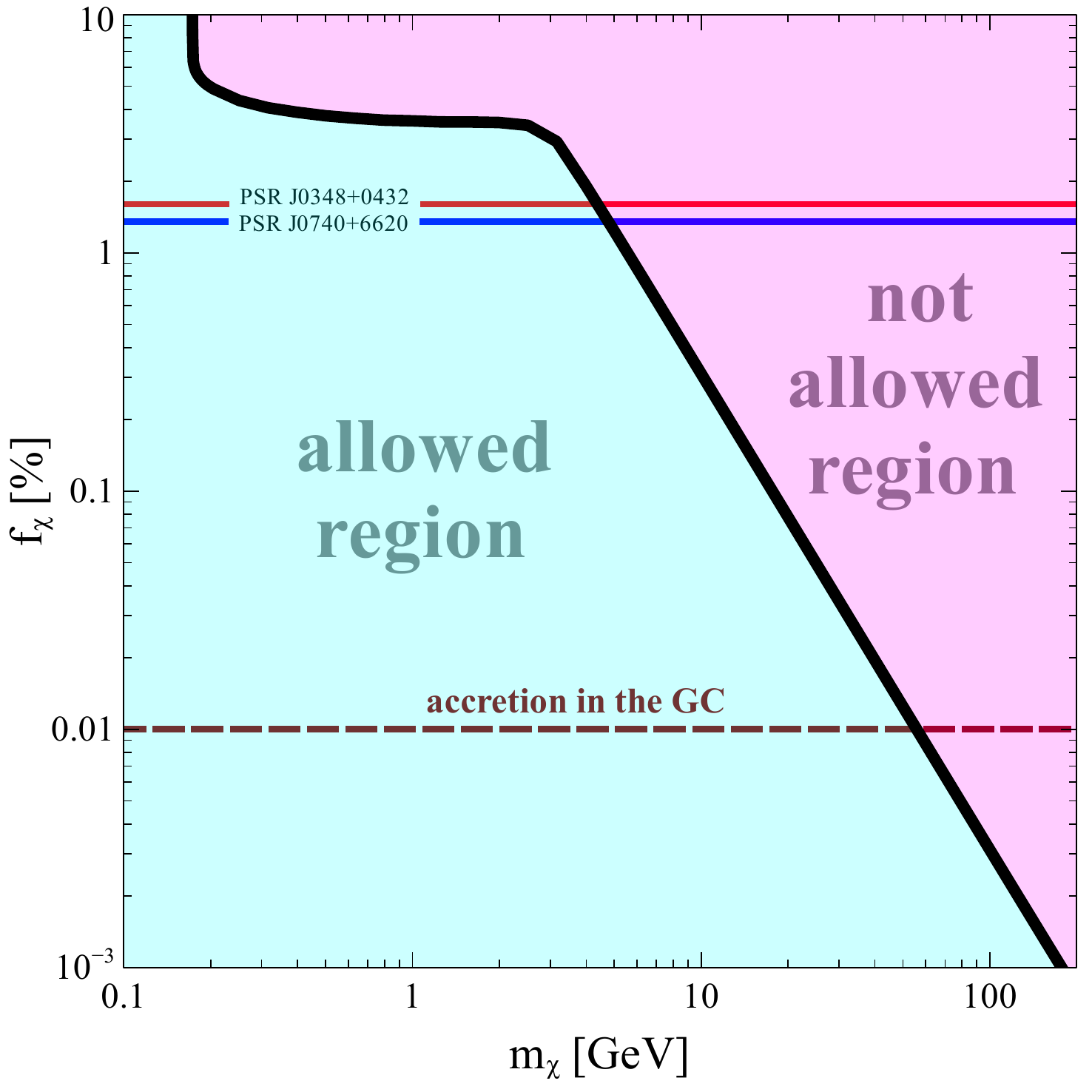}  
\caption{The critical fraction of DM $f_\chi^c$ vs. its particle mass $m_\chi$. The red and blue lines correspond to the DM fraction $f^*_{\chi}$ in the surrounding medium around the heaviest known pulsars (see the text for details). The pink area above the black curve represents unphysical region with $M_{max}<2\,M_\odot$, while in the cyan region $M_{max}>2\,M_\odot$. Rough estimation for the fraction of accreted DM into the NSs in the most central region of the Galaxy is depicted as the dashed dark red line.}
\label{fig4}
\end{figure}
Below for each value of $m_\chi$ we find a critical fraction of DM $f_\chi^c$, which yields to $M_{max}=2~M_\odot$. Fig. \ref{fig4} shows how this quantity (black curve) varies with $m_\chi$. Any point above this curve corresponds to the unphysical region with $M_{max}<2\,M_\odot$, and, consequently, is inconsistent with the two solar mass observational limit \citep{2019arXiv190406759C, 2013Sci...340..448A}. All the points below this curve agree with the observational condition $M_{max}>2\,M_\odot$. Note, that for 5 GeV $<m_\chi<$ 1 TeV, the critical fraction of DM is well fitted by the power law: $f_\chi^c\sim m_\chi^{-1.92}$. Hence, knowing $f_\chi^c$ as a function of $m_\chi$ we can set a constraint on the DM particle mass.

\paragraph*{}
First, we estimate the DM fraction in the stellar surrounding, which depends on the distance from the Galaxy centre (GC) $d$. We denote this quantity $f_\chi^{*}$, and compute it by using two DM density profiles, i.e. the Navarro-Frenk-White (NFW)  \citep{1996ApJ...462..563N} and the Einasto \cite{Einasto1965} spatial mass distributions of DM in the Milky Way. Recent work \cite{2014MNRAS.441.3359D} has shown a better agreement of N-body simulations with the Einasto profile. In particular, we are interested in the local fraction of DM $f_\chi^{*}$ in regions near the heaviest known pulsars, specifically,  PSR J0348+0432 with $M=2.01\pm 0.04M_\odot$ \citep{2013Sci...340..448A} and PSR J0740+6620 with $M=2.14^{+0.10}_{-0.09}M_\odot$ \citep{2019arXiv190406759C}. Their distances to the GC are $9.9$ kpc and $8.6$ kpc, respectively. 

The shape of the NFW profile
\begin{eqnarray}
\label{X}
\rho_\chi(d) = \rho_c\cdot\frac{d_c}{d}\cdot\left(1+\frac{d}{d_c}\right)^{-2}
\end{eqnarray}
is controlled by the characteristic density $\rho_c=5.22\pm0.46\,10^7\,{M_\odot}{\rm kpc^{-3}}$ and characteristic length $d_c=8.1 \pm 0.7~{\rm kpc}$ determined in Ref. \cite{2019MNRAS.tmp.1610L} directly from the observational data analysis.

The Einasto profile is defined as
\begin{eqnarray}
\label{XI}
\rho_\chi(d) = \rho_{-2} e^{-\frac{2}{\alpha} \left[ \right( \frac{d}{r_{-2}}\left)^{\alpha} -1  \right]}
\end{eqnarray}
with $\rho_{-2}$ being the DM density at the distance $r_{-2}$, where the logarithmic gradient $\frac{d~{\rm ln}\rho}{d~{\rm ln}r}= -2 $. The curvature of this profile is defined by the parameter $\alpha$ that ranges between 0.15 and 0.19 for the Milky Way type galaxies \cite{2008MNRAS.390L..64D}. Following Ref. \cite{2010PhRvD..81l3521D} we allow small corrections for the effects of baryonic contraction and adopt the values $r_{-2}=16~ ~{\rm kpc}$ and $\alpha=0.19$. The value of $\rho_{-2}$ is found with the help of Eq. (\ref{XI}) using the fact that the local DM density in the Sun's surrounding is estimated to be $\rho_\chi(d_{\odot})\simeq 0.420 ^{+0.019}_{-0.021}  ~{\rm GeV/cm^{3}}$ \cite{2015JCAP...12..001P} ($d_{\odot}$ is a distance to the Sun from the GC).

Two heaviest known pulsars are located far enough from the Galaxy bulge (with a size $d_b\simeq1.9$ kpc) \citep{2013PASJ...65..118S}. Therefore, with high accuracy, we can consider only the contribution of the stellar disc to the BM density profile near these pulsars:
\begin{eqnarray}
\label{XII}
\rho_B(d)=\rho_{dc} e^{-\frac{d}{d_{dc}}}.
\end{eqnarray}
Here $\rho_{dc}=15.0\;{M_\odot}{\rm pc^{-3}}$ and $d_{dc}=3.0$ kpc  \citep{2013PASJ...65..118S}.

\paragraph*{}
We estimate the local fraction of DM in the Milky Way as the ratio of DM mass density to the total mass density of DM and BM. Thus, using Eq. (\ref{X}) for the NFW distribution and Eq. (\ref{XII}), we find $f_\chi^*=1.6\pm0.4~\%$ near PSR J0348+0432 and $f_\chi^*=1.35\pm0.35~\%$ near PSR J0740+6620. On Fig. \ref{fig4} these values are depicted as red, green and blue lines, respectively. For the Einasto profile given by Eq. (\ref{XI}) the estimated fractions $f_\chi^*$ equal to $ 1.35\pm0.05~\%$, $ 1.12\pm0.049~\%$ near PSR J0348+0432 and PSR J0740+6620, correspondingly. We want to emphasise that such a straightforward analysis gives us the upper estimate of the fraction of DM inside the NSs. In the next section we discuss the relevance of this estimate in comparison to the accretion and thermalisation rates of DM particles inside the NSs.

%%%%%%%%%%%%%%%%%%%%%%%%%%%%%%%%%%%%%%%%%%%%%%%%%%%%%%%%%%%
\section{Dark matter accumulation regimes}
\label{sec-5}
 
The amount of DM in NS is expected to vary depending on the evolution history and position of the star. Thus, the environment from which it originates and its distance from the GC define properties of the progenitor and DM accretion rate during the whole lifetime of the star, including (i) progenitor, (ii) main sequence (MS) star, (iii) supernova explosion with formation of a proto-NS and (iv) equilibrated NS phases. Below we consider accumulation of DM at each of these stages.

i. DM causes a gravitational effect in the form of tidal energy that promotes a star formation, especially in the innermost region of the Galaxy. During the star formation stage the initial mixture of DM and BM contracting to form the progenitor star. Trapped DM undergoes scattering processes with baryons leading to its thermalisation, and energy loss. This process becomes more significant in the later stages of evolution when the star starts to produce heavier elements.

ii. An accretion of DM into the MS stars has been extensively studied in the literature, e.g. in Refs. \cite{2009MNRAS.394...82S, 2019ApJ...880L..25L}. From this stage, the DM accretion rate increases due to the fact that the star creates a stronger gravitational potential. Depending on the distance from the GC the accreted mass by the star can range from $ \sim 10^{-5}M_{\odot}$ to $10^{-9} M_{\odot}$ at the central region of the Galaxy.

iii. Gravitational collapse with a following supernova explosion marks the end of a massive star that has exhausted all its fuel. The proto-NS formed during a supernova is characterized by the dynamical instabilities, inhomogeneities of matter in its interior and high fraction of heavy elements that increase the interaction rate with the DM particles.

The newly-born NS will be surrounded by the dense cloud of DM particles with the temperature and radius that corresponds to the last stage of MS star evolution, i.e. a star with a silicone core. Therefore the accretion onto the proto-NS and DM density in surrounding media might be affected by the previous stage \cite{2010PhRvD..82f3531K}. 

In addition, a significant amount of DM can be produced during the supernova explosion and mostly remain trapped inside the star \cite{2018arXiv180303266N}.

iv. As it was shown in Refs. \cite{2008PhRvD..77b3006K, 2010PhRvD..82f3531K}, due to the spherically symmetric accretion during the phase of equilibrated NS an amount of accreted DM mass can be found as 
\begin{eqnarray}
\label{XIII}
\hspace*{-0.2cm} M_{acc} \approx 10^{-14} \hspace*{-0.05cm} \left( \frac{\rho_\chi}{0.3~ \frac{GeV}{cm^{3}}} \right) \hspace*{-0.05cm} \left( \frac{\sigma_{\chi n}} {10^{-45} cm^{2}} \right) \hspace*{-0.05cm} \left( \frac{t}{Gyr} \right) \hspace*{-0.05cm} M_{\odot},
\end{eqnarray}
where $t$, $\rho_\chi$ are the accretion time and the local DM density, correspondingly.
Thus, considering a typical value for the elastic cross section $\sigma_{\chi n}$ between DM and nucleons of the order of $10^{-45} cm^{2}$, the mean free path of DM particle inside the NS is a couple of km. Obviously, for the larger value of cross section, the fraction of the captured DM particles saturates even faster to 1.

For obtained DM densities around the two heaviest pulsars calculated using the NFW (see Eq. \ref{XI}) and the Einasto  (see Eq. \ref{XI}) profiles the amount of accreted matter is  $ \sim 10^{-13} - 10^{-14} M_{\odot}$. As the NFW profile has a divergent inner density profile of $\rho_\chi(d) \propto d^{-1}$ (see Eq. \ref{XI}), for description of the central parts of the Galaxy we use only the Einasto profile. An accreted mass in the most central Galaxy region can range between $10^{-5} M_\odot - 10^{-8} M_\odot$ depending on the value of $\alpha$ parameter \cite{arXiv:1904.13060}.

Furthermore, in the most central parts of the Galaxy clumps of DM that can reach masses ranging from $10^{-6} M_\odot$ to $10^{2} M_\odot$ \cite{2006PhRvL..97c1301P} could be accreted on NSs. Such a mechanism may substantially increase the DM fraction inside the compact stars.

As was outlined in this section, an accretion of DM onto the MS star and NS are the most studied ones and give a comparable efficiency. A former stage (ii) characterises with a larger mass and radius, and, therefore, bigger capturing surface. However, higher density of NSs favours DM capturing after a single crossing \cite{2010PhRvD..82f3531K}.
Considering only these two stages the estimated fraction of DM in the GC can reach up to $10^{-4} M_\odot$. Note, that this value does not include possible presence of DM in the proto-star cloud, production of DM particles during the supernova explosion and accretion of DM clumps.

On the Fig. \ref{fig4} the above estimation is shown as the dashed dark red line. Its intersection with the black curve makes us to conclude that measurements of a $2 M_\odot$ NS in the GC will impose an upper constraint on the mass of DM particles of $\sim$ 60 GeV. More precise modelling of DM accumulation inside the NSs will put more tight constraints on the mass of DM particles.

%%%%%%%%%%%%%%%%%%%%%%%%%%%%%%%%%%%%%%%%%%%%%%%%%%%%%%%%%%%
\section{Searches for the dark matter admixed neutron stars}
\label{sec-6}

As it was discussed in Sec. \ref{sec-3} and in Ref. \citep{2018arXiv180303266N}, DM can either create an extended halo around the NS or condensate in the core of a star. In both these scenarios accumulated DM can affect the thermodynamic properties of matter, its tidal polarisability and the GW signal during the merger \citep{2019PhRvD.100d4049B, 2018PhLB..781..607E}. In case of dark halos that may  extend for more than 100 km we can expect changes in the dynamics of the merger. Thus, coalescence of the DM halos would start much earlier before the merger of the baryonic components of the NSs creating a viscous envelope around them. To understand dynamics and gravitational radiation emitted during the coalescence of such a system full 3D numerical simulations are required. For the ongoing and future generations of GW detectors such prominent effect on the merger dynamic opens a new possibility of detection DM admixed compact stars.

On the other hand, study of the thermal evolution of NSs is an another viable indirect method to test the presence of DM in star's interior. Thus, in Ref. \cite{2016PhRvD..93f5044S} it is shown how the emission of DM particles by weakly magnetized NSs during their early and intermediate evolution stages alters observable surface temperatures, while another study reveals a late-stage heating caused by DM particle annihilation inside the NS core \cite{2019PhLB..795..484H}. 

The upcoming radio and X-ray surveys are expected to significantly enrich the number of detected NSs and to provide a more accurate information about the mass of NSs, especially in the central region of Milky Way, where the fraction of DM is higher. The number of pulsars and magnetars in the most central part of the Galaxy within $70\,pc$ from its center is likely to be high \citep{2018ASPC..517..793B}. So far, only six pulsars in the central region, including a transient magnetar J1745-2900 \cite{2018ApJ...866..160P} are known. These searches are hampered by an extreme scattering of pulsar radio emission as a result of large electron density along the line of sight.
 
In this context, the following astronomical projects are very promising: {\it radio telescopes} -- the Karoo Array Telescope (MeerKAT) \citep{2018arXiv180307424B}, the Square Kilometer Array (SKA) \citep{2015aska.confE..43W},  and the Next Generation Very Large Array (ngVLA) \citep{2018IAUS..336..426M}; {\it space telescopes} --	the Neutron Star Interior Composition Explorer Mission (NICER) \citep{2019arXiv190407012W}, the Advanced Telescope for High Energy Astrophysics (ATHENA) \citep{2018arXiv180709080C}, the enhanced X-ray Timing and Polarimetry mission (eXTP) \citep{2019SCPMA..6229506I}, and the Spectroscopic Time-Resolving Observatory for Broadband Energy X-rays (STROBE-X) \citep{2019arXiv190407012W}. They are expected to provide simultaneous measurements of mass and radius of the NSs, their moment of inertia and advance for 10\% the time precision for double pulsars. An improvement in mass determination will answer to the question whether or not we see a mass reduction of the NSs in the close proximity to the GC.

%%%%%%%%%%%%%%%%%%%%%%%%%%%%%%%%%%%%%%%%%%%%%%%%%%%%%%%%%%% 
\section{Summary and Conclusions}
\label{sec-7}

Using the observational fact of the existence of the two heaviest known NSs (i.e., PSR J0348+0432, PSR J0740+6620) with the masses exceeding the two solar ones, we present a new allowable range of masses of DM particles and their fractions inside the star. Our analysis is based on the ability of NSs to accumulate a sizeable amount of ADM, which can significantly reduce the mass of the host NS. We also demonstrate that DM lighter than 0.2 GeV can create an extended halo around the NS leading not to decrease but to increase of the NS total (gravitational) mass. By using\- recent results on the distribution of DM and BM in the Milky Way we calculated the DM fraction in the surrounding medium around the heaviest known pulsars.

Furthermore, we discussed the main stages of star evolution from the progenitor to the NS and their influence on the capturing rate of DM. Keeping in mind our limited understanding of the amount of DM in the proto-star cloud, impact of a supernova explosion and the proto-NS stages, NS in the most central parts of the Galaxy can maintain about 0.01\% of DM from the total mass of the star. Based on this estimation we argue that measurements of a $2 M_\odot$ NS in the center of the Milky Way will constrain the upper mass of the particles of ADM below 60 GeV.

These results are of big interest for the forthcoming and ongoing projects, which are planned to increase the total number of observed compact objects by a factor $\sim 10$ and to provide a better determination of their masses and radii \citep{2018arXiv180307424B,2015aska.confE..43W,2018IAUS..336..426M,2019arXiv190407012W,2018arXiv180709080C,
2019SCPMA..6229506I}.

%%%%%%%%%%%%%%%%%%%%%%%%%%%%%%%%%%%%%%%%%%%%%%%%%%%%%%%%%%%
\section{Acknowledgments}
\label{sec-8}

The authors are thankful to A. Del Popolo, M. Le Delliou, C. Miller, A. Rudakovskyi and D. Savchenko for fruitful discussions and valuable comments. This  work  was  supported by the Funda\c c\~ao para a Ci\^encia e Tecnologia (FCT), Portugal, under the projects UID/04564/2020, UIDB/00099/2020, CENTRO-01-0145-FEDER-000014 through CENTRO2020 program, and  POCI-01-0145-FEDER-029912 with financial support from POCI, in its FEDER component, and by the FCT/MCTES budget through national funds (OE). O.I. acknowledges the Universidad de Salamanca, Spain, for the support within the project SA083P17 launched by the Regional Government of Castilla y Leon/the European Regional Development Fund and for the National Academy of Sciences of Ukraine for the partial support by the project No. 0118U003197.

%%%%%%%%%%%%%%%%%%%%%%%%%%%%%%%%%%%%%%%%%%%%%%%%%%%%%%%%%%%%
\section*{APPENDIX}
\label{sec-app}
\setcounter{section}{1}
\renewcommand{\thesection}{\Alph{section}}
\setcounter{equation}{0}
\renewcommand{\theequation}{\thesection.\arabic{equation}}

Here we derive the system of relativistic hydrostatic equations Eqs. (\ref{I}) in a manner closely resembling the one of Ref. \cite{1939PhRv...55..374O}. We start from the Einstein's filed equations in the absence of the cosmological constant
\begin{eqnarray}
\label{A1}
-8\pi T^{\mu\nu}=R^{\mu\nu}-\frac{1}{2}Rg^{\mu\nu}.
\end{eqnarray}
Due to the negligibly weak non gravitational interaction between BM and DM the only non vanishing components of the stress-energy tensor in the perfect fluid approximation can be written as
\begin{eqnarray}
\label{A2}
T^{11}=T^{22}=T^{33}=p_B+p_D,\quad T^{44}=\epsilon_B+\epsilon_D.
\end{eqnarray}
In the spherically symmetric case metrics attains the well known form
\begin{eqnarray}
\label{A3}
ds^2=-e^\lambda dr^2-r^2d\theta^2-r^2\sin^2\theta d\phi^2+e^\nu dt^2.
\end{eqnarray}
Further, Eqs. (\ref{A1}) reduce to three independent equations for metric functions $\lambda$ and $\nu$ together with profile of the total energy density $\epsilon_B+\epsilon_D$, connected to the total pressure $p_B+p_D$ through EoSs of BM and DM. Introducing a new variable 
\begin{eqnarray}
\label{A4}
M=\frac{2}{R}\left(1-e^{-\lambda}\right)
\end{eqnarray}
these three equations turn to
\begin{eqnarray}
\label{A5}
\frac{dM}{dr}&=&4\pi r^2(\epsilon_B+\epsilon_D),\\
\label{A6}
\frac{d\nu}{dr}&=&\frac{2M+8\pi r^3p}{r^2(1-2M/r)},\\
\label{A7}
\sum_{j=B,D}\frac{dp_j}{dr}&=&-
\sum_{j=B,D}\frac{p_j+\epsilon_j}{2}\frac{d\nu}{dr}.
\end{eqnarray}
To proceed further we utilize the thermodynamic identities $n_j=\frac{\partial p_j}{\partial\mu_j}$ and $\epsilon_j=\mu_j n_j-p_j$, which lead to $\frac{dp_j}{dr}=n_j\frac{d\mu_j}{dr}$ and $p_j+\epsilon_j=n_j\mu_j$, respectively. Thus, Eq. (\ref{A7}) becomes
\begin{eqnarray}
\label{A8}
\sum_{j=B,D}n_j\mu_j
\left[\frac{d\ln\mu_j}{dr}+\frac{1}{2}\frac{d\nu}{dr}\right]=0.
\end{eqnarray}
In the equilibrium case chemical potential in the local Lorentz frame scales as $\mu_j\sqrt{g_{44}}=const$ leading to $\frac{d\ln\mu_B}{dr}=\frac{d\ln\mu_D}{dr}$. Therefore, square bracket in the previous relation is the same for $j=B$ and $j=D$ and, thus, can be taken out of the sum. Since $n_B\mu_B$ and $n_D\mu_D$ are positively defined, then Eq. (\ref{A8}) simplifies to
\begin{eqnarray}
\label{A9}
\frac{d\ln\mu_j}{dr}+\frac{1}{2}\frac{d\nu}{dr}=0,
\end{eqnarray}
which is valid for both BM and DM. This result yields
\begin{eqnarray}
\label{A10}
\frac{dp_j}{dr}=n_j\mu_j\frac{d\ln\mu_j}{dr}=
-\frac{n_j\mu_j}{2}\frac{d\nu}{dr}=
-\frac{p_j+\epsilon_j}{2}\frac{d\nu}{dr}.\quad
\end{eqnarray}
Combining this relation with Eq. (\ref{A6}) we end up with Eq. (\ref{I}).

%%%%%%%%%%%%%%%%%%%%%%%%%%%%%%%%%%%%%%%%%%%%%%%%%%%%%%%%%%%

%%%%%%%%%%%%%%%%%%%%%%%%%%%%%%%%%%%%%%%%%%%%%%%%%%%%%%%-
 
\end{document}